\documentclass[twocolumn,showpacs,preprintnumbers,amsmath,amssymb,aps,prb,10pt]{revtex4-1}

\usepackage{graphicx}
\usepackage{subfigure}
\usepackage{bm}
\usepackage{color}
\newcommand{\bi}{\begin{itemize}}
\newcommand{\ei}{\end{itemize}}
\newcommand{\be}{\begin{eqnarray}}
\newcommand{\ee}{\end{eqnarray}}
\newcommand{\nn}{\nonumber}

\newcommand{\re}{\color{red}}
\newcommand{\bl}{\color{blue}}

\begin{document}

\title{Bench tests for microscopic theory of Raman scattering in powders of disordered nonpolar crystals: nanodiamonds and {\rm Si} nanoparticles}

\author{Andrey G. Yashenkin$^{1,2}$}

\author{Oleg I. Utesov$^{1,2,3}$}
\email{utiosov@gmail.com}

\author{Sergei V. Koniakhin$^{4,5}$}
\email{kon@mail.ioffe.ru}

\affiliation{$^1$Petersburg Nuclear Physics Institute NRC ``Kurchatov Institute'', Gatchina 188300, Russia}
\affiliation{$^2$Department of Physics, St. Petersburg State University, St.Petersburg 199034, Russia}
\affiliation{$^3$St. Petersburg School of Physics, Mathematics, and Computer Science, HSE University, 190008 St. Petersburg, Russia}
\affiliation{$^4$Institute Pascal, PHOTON-N2, University Clermont Auvergne, CNRS, 4 Avenue Blaise Pascal, Aubi\`{e}re Cedex 63178, France}
\affiliation{$^5$St. Petersburg Academic University - Nanotechnology Research and Education Centre of the Russian Academy of Sciences, St.Petersburg 194021, Russia}

%\author{Andrey G. Yashenkin}
%\affiliation{Petersburg Nuclear Physics Institute NRC ``Kurchatov Institute'', Gatchina 188300, Russia}
%\affiliation{Department of Physics, St. Petersburg State University, St.Petersburg 199034, Russia}

%\author{Oleg I. Utesov}
%\email{utiosov@gmail.com}
%\affiliation{Petersburg Nuclear Physics Institute NRC ``Kurchatov Institute'', Gatchina 188300, Russia}
%\affiliation{Department of Physics, St. Petersburg State University, St.Petersburg 199034, Russia}

%\author{Sergei V. Koniakhin}
%\email{kon@mail.ioffe.ru}
%\affiliation{Institute Pascal, PHOTON-N2, University Clermont Auvergne, CNRS, 4 Avenue Blaise Pascal, Aubi\`{e}re Cedex 63178, France}
%\affiliation{St. Petersburg Academic University - Nanotechnology Research and Education Centre of the Russian Academy of Sciences, St.Petersburg 194021, Russia}

\date{\today}

\begin{abstract}

Recent Raman data on diamond and crystalline {\rm Si} nanopowders are analyzed with the use of a microscopic theory of Raman scattering in ensembles of disordered nonpolar nanocrystals. The large width of the Raman peak in nanoparticles as compared to the corresponding peak in bulk materials and the peak inverse dependence on the particle size observed experimentally are explained within the framework of the theory. It is shown that this theory is capable to extract confidently from the Raman data four important microscopic characteristics of the nanopowder including the mean particle size, the variance of the particle size distribution function,  the strength of intrinsic disorder in the particle, and the effective faceting number which parameterizes the particle shape.

\end{abstract}

\maketitle
\section{Introduction}

Multifarious nanoparticles, assembled in ordered arrays of quantum dots, photonic crystals, etc., \cite{roh2020optically} or existing in random formations of powders or liquid suspensions~\cite{geiregat2019bright} are apparently the most intensively studied objects in modern physics and chemistry. Close attention to these entities is fuelled by their possible scientific and industrial applications~\cite{veldhorst2014addressable,kidalov2009thermal,xia2013nanoparticles}. Nonpolar nanocrystals, including diamond-like and semiconducting ones are very promising candidates for technological utilizations~\cite{behler2009nanodiamond,pichot2019nanodiamond}. Among them, the special role is played by carbon nanostructures due to their good bio-compatibility, optical, electronic, and mechanical properties. Various families of nanodiamonds (detonation, laser, HPHT synthesis) inherit the outstanding properties of bulk diamonds and thus are of special interest. A great amount of experimental methods, namely X-ray diffraction, dynamical light scattering, atomic force microscopy, Raman scattering, etc., are used in order to investigate these materials~\cite{pecora2000,chu2007laser,mourdikoudis2018characterization,koniakhin2015molecular,koniakhin2018ultracentrifugation,hassan2015making,segets2016analysis}.

The Raman spectroscopy plays an essential role in the characterization of carbon nanomaterials. It is a precise nondestructive instrument to observe the peculiarities of collective excitations (usually, optical phonons) in various materials~\cite{cardona2006light,kumar2012raman}. For the carbon nanostructures this method gives even more information than for other materials: distinction of sp$^2$/sp$^3$/amorphous phases \cite{ferrari2004raman}, defectness and the number of layers in graphene, graphitization degree, and the size of nanodiamonds. And {\it vice versa}, among the materials studied by means of the Raman spectroscopy significant fraction belongs to the carbon nanostructured materials including various types of diamond nanoparticles.

%\textit{In present study we focus on Raman spectra of nanodamonds, namely on the peculiarities of the shape of diamond crystalline peak. As it was mentioned above, along with probing phase composition by Raman spectroscopy, the size of diamond core can be estimated using the rough Phonon Confinement Model or extracted with better accuracy based on advanced DMM-BPM or EKFG theories. All these three theories explain the Raman peak downshift by the size quantization of optical phonons in nanoparticles. Here, DMM-BPM and EKFG theories will be used as a grounds to calculate the optical phonon lifetime in nanocrytallites and thus the broadening of Raman peak, the quantity taken only phenomenologically for nowdays.}

Reliable analysis of the shape and  position of the diamond crystalline Raman peak will equip us with detailed information about the nanoparticle ensemble. The phenomenon of finite-size quantization of the momentum in particles results in a size-dependent shift of the peak as compared to bulk materials~\cite{meilakhs2016new}. The full spectrum of vibrational modes which form this peak depends on the particle shape. Recently, we have developed two theories, which we called DMM-BPM \cite{ourDMM} and EKFG \cite{ourEKFG} approaches,  capable to evaluate the Raman data more precisely than the previously used phonon confinement model (PCM)~\cite{richter1981one,campbell1986effects,adu2005confined,faraci2006modified}. Notice that the current efforts of the community aim mainly to incorporate the unrealistically fine effects into the PCM~\cite{osswald2009phonon,korepanov2017quantum,korepanov2017,korepanov2017quantum,zi1997comparison,faraci2006modified,ke2011effect}; however, some alternative approaches (e.g., the local-mode model~\cite{gao2019determination}) are also proposed.

Both our theories capture the principal features of the optical phonon spectra in diamond-like nanoparticles, including (i)~existence of ``Raman active'' (contributing to Raman) and ``Raman silent'' (not contributing) eigenmodes and (ii)~presence of the first (degenerate) level which provides the majority (about 2/3 for a diamond) of the total spectral weight being separated from the rest of  spectrum by a huge gap. The spectrum forms (iii)~several ``bands'', each of them could be treated as a (quasi)continuum. (iv)~The structure of these bands depends on the particle shape.
These theories successfully explain recent experimental data on nanodiamonds and semiconducting nanocrystals~\cite{ourDMM, ourEKFG}; however, they require the fitting parameter $\Gamma$ for the optical phonon lines broadening, introduced by hands.

The disadvantage of all mentioned (DMM, EKFG, local-mode, PCM) approaches is the lack of microscopic mechanism to provide the parameter $\Gamma$ finite. No microscopic explanation of the origin of phonon damping exists in literature, as well (see phenomenological analysis of experiments in Refs.~\cite{yoshikawa1993raman,yoshikawa1995raman,chaigneau2012laser}). This gap has been filled in by our papers ~\cite{our3,our4,ourK}, where the DMM-BPM and EKFG theories have been used as  starting points to calculate the optical phonon lifetimes $\Gamma$ in nanocrytallites and thus to broaden the Raman peak. The origin for these phonon damping have been attributed to the intrinsic (including surface) disorder always existing in nanoparticles. We investigated (both analytically and numerically) the disorder taken in the form of random atomic masses in Refs.~\cite{our3,our4} and stemming from defective interatomic bonds in Ref.~\cite{ourK}, in line with more involved types of disorder (random smooth disorder, surface corrugations, surface amorphization) treated numerically. While the above analysis have been concentrated on weakly disordered {\it nanocrystals}, later on it has been extended onto the Raman spectra of {\it amorphous} silicon~\cite{ranipredicting}.

This paper is aimed to promote the theory developed in Refs.~\cite{ourDMM,ourEKFG,our3,our4,ourK} as a regular powerful method of interpreting the Raman spectra of nanopowders of nonpolar crystals. The best way to do it is to re-examine existing experiments, extracting from the data precise and detailed information.
The results of this analysis are as follows. First, we demonstrate that the well-known but unexplained phenomenon of strong broadening of the main Raman peak in nanoparticles as compared to the corresponding bulk material could be easily explained within the framework of our theory: the typical Raman peak widths are achievable at realistic concentrations of experimentally relevant types of disorder. Second, we find that the empirical inverse dependence of the Raman peak width on the particle size $L$ reported by Yoshikawa with co-authors in Ref.~\cite{yoshikawa1993raman} could be attributed to our formula for the optical phonon damping in disordered nanoparticles, $\Gamma \propto 1/L$, evaluated for the regime of overlapped phonon levels. Moreover, analyzing the experimental conditions of Ref.~\cite{yoshikawa1993raman} we observe, that they correspond precisely to this regime.
Third, re-examining three sets of experimental data (two on nanodiamonds and one on {\rm Si} nanocrystals) we were capable to extract from these data such important
characteristics of the nanopowders as (i) the mean particle size  (ii) the variance of size distribution function, (iii) the  nanoparticle shape (facet number), and (iv) the strength of disorder; for {\rm Si} data, we also improve the spectral parameters of optical phonons.

Importantly, the methods developed in Refs.~\cite{ourDMM,ourEKFG,our3,our4,ourK} could be applied for various crystalline nanoparticles including diamonds, Si, Ge nanoparticles, GaAs, and other quantum dots either embedded to the matrix or not. To do this, only an optical phonon dispersion, atomic masses, interatomic bond rigidities, and the parameters of lattice defects should be modified.

The paper is organized as follows. In Section~\ref{II} we briefly scketch the DMM-BPM and the EKFG theories. Section~\ref{III} outlines the optical phonon line damping.
In Section~\ref{IV} we check numerically the analytical approaches of previous two sections. Section~\ref{V} is concentrated on the analysis of experimental data with the use of theories developed earlier. The last Section~\ref{VI} is reserved for the discussion of final results and for  concluding remarks.

\section{Two theories of Raman spectra in nanoparticles } \label{II}

The DMM-BPM~\cite{ourDMM} theory has a discrete atomistic character. It is built upon the direct evaluation of  $3N \times 3N$ dynamical matrix \cite{born1954dynamical}, which allows to determine all vibrational modes (i.e., their eigenfrequencies and eigenfunctions) for a nanoparticle of given shape  containing $N$ atoms:
 \begin{equation}\label{DMM}
M  \, \omega^2 r_{i, \alpha} = \sum_{j=1}^{N} \sum_{\beta =x,y,z} \frac{\partial^2 \Phi}{ \partial r_{j, \alpha} \partial \, r_{j, \beta}} r_{j, \beta},
\end{equation}
where ${\bf r}_{i}$ and $M$ are the instant displacement of $i$-th atom from the equilibrium position and its mass,  $\alpha,\beta = x,y,z$, $\omega$ is the frequency, and $\Phi$ is the total particle energy. In order to specify the function $\Phi$ we used the Keating model~\cite{keating1966effect}; however, any (mechanistic) model of a crystal which deals with its potential energy as a function of atomic displacements is acceptable. In particular, the engagement of the Keating model allows to express the constants $\omega_0$ and $F$  in the spectrum of long wavelength optical phonons
\begin{equation} \label{omega1}
\omega (q_n) \equiv \omega_n \approx \omega_0 \,[ 1 - F (q_n a_0)^2]
\end{equation}
via the Keating spectral parameters $A$ and $B$
\begin{equation} \label{omega2}
\omega_{\bf q} = A + B \, {\rm cos} \, (q a_0 /2),
\end{equation}
 providing $\omega_0 = A + B$ and $F=B/8(A+B)$, the former should be traced back to the quantum chemical quantities $\alpha_0$ (bond rigidity with respect to stretching) and $\beta_0$ (valence angle bending) as follows:
\begin{equation}
\omega_{0}^{2} = \frac{8}{M} (\alpha_0 + \beta_0).
\end{equation}
Here $\omega_n$ is the frequency of the optical phonon mode with quantum number $n$, $\omega_0$ is the maximal frequency of this mode, $F$ is the spectrum flatness parameter, $q_n$ is the discrete phonon quasimomentum in a finite-size particle, and $a_0$ is the lattice constant.

The eigenfunctions/eigenvalues of Eq.~\eqref{DMM} could be converted into the Raman spectra within the framework of the  bond polarization model~\cite{martin1984raman,snoke1993bond}, provided that the polarization of a crystal occurs exclusively due to atomic displacements (which is the case for nonpolar crystals)
\begin{equation} \label{P}
P_{\alpha \beta} (n) = \sum_{i=1}^{N} \sum_{\gamma} \, M_{\,\, i}^{ \alpha \beta \gamma} \, r_{i, \gamma} (n).
\end{equation}
Here $P_{\alpha \beta} (n)$ is the polarization tensor for $n$-th mode, and $M_{\,\,i}^{ \alpha \beta \gamma}$ is certain (known) combination of atomic radius vectors and material constants which could be expressed via the microscopic parameters of the theory~\cite{ourDMM}.
The resulting Raman spectrum for the ensemble of  particles of a given shape (parameterized say by the number of faces of equivalent regular polytopes $p$), which have the sizes $L$, obtains the form
\begin{equation} \label{I1}
I (\omega, L, p) \propto \sum_n I_{n}(L,p)  \frac{\Gamma_n (L, p)}{(\omega - \omega_n (L,p))^2 + \Gamma_{n}^{2} (L,p)}.
\end{equation}
Here $\Gamma_n (L,p)$ are the phonon damping parameters~\cite{our3,our4,ourK} specified in the next section (in the initial theory of Refs.~\cite{ourDMM,ourEKFG} they were introduced empirically).
The quantities $I_{n} (L,p)$  entering Eq.~\eqref{I1} originate from the matrix elements of the effective photon-phonon interaction and contain the phonon eigenfunctions averaged over the particle volume and then squared. Due to their symmetry properties some phonon modes do not enter the final result being averaged to zero for particles of particular shapes. This results in the concept of ``Raman-active'' and ``Raman-silent'' bands in the Raman spectrum (see, e.g., Refs~\cite{ourEKFG,our3}). As a result, the Raman spectrum is strongly affected by the particle shape.

The last step of spectral calculations for a nanopowder is the averaging of Eq.~\eqref{I1} over the size (or, sometimes, size and shape) distribution function $f(L)$ which yields $I (\omega, p) = \sum_{L} I (\omega, L,p) f(L)$.

The Raman spectra calculated this way excellently fit the experimental data on nanopowders of nonpolar crystals without adjusting parameters.
However, it requires quite a long time for computer to manipulate with $3N \times 3N$ matrices, so size of particles accessible for DMM-BPM calculations is presently limited by 5-6 {\rm nm}. In order to proceed with larger particles we developed the continuous EKFG approach described below.

More specifically, it has been demonstrated in Ref.~\cite{ourEKFG} that the long wavelength limit of the (discrete isotropic)  DMM problem \eqref{DMM} for optical phonons is governed by the continuous Klein-Fock-Gordon equation in the Euclidean space (EKFG) with Dirichlet boundary conditions:
\begin{equation}\label{EKFG}
(\partial^2_t + C_1 \Delta + C_2 ) \, Y = 0, \quad Y|_{\partial \Omega} = 0.
\end{equation}
Here $C_{1,2}$ are the positive constants which have one-to-one correspondence to the parameters of DMM theory (and, therefore, to the quantum chemical quantities), $Y$ are the (scalar) eigenfunctions of the considered Sturm-Liouville problem for EKFG equation \eqref{EKFG}, $\partial \, \Omega$ is the particle boundary. This continuous problem could be solved much faster and for larger particles than the original discrete DMM one given by Eq.~\eqref{DMM}. Its approximate character
manifests itself in deviations of the solution at frequencies lying relatively far away from the position of maximum of the main Raman peak; however, our analysis in
Refs.~\cite{ourEKFG,our3,our4} demonstrates that for particles with $L \leq 10-20 \, {\rm nm}$ the difference in the fit of the main Raman peak by means of DMM-BPM and EKFG approaches is almost indistinguishable.

The EKFG apparatus accompanied by the properly modified continuous version of the BPM theory \cite{ourEKFG} provides us with the second tool allowing to interpret microscopically the Raman spectra of nanopowders of nonpolar crystals.

At the end of this section, let us present here one more useful formula which utilizes the scaling properties of Eq.~\eqref{EKFG}. It allows to build up rapidly and without tedious calculations the Raman spectrum of a nanopowder with a given size distribution function starting from the spectrum calculated for a single particle with size $L$:
\begin{equation}\label{scaling}
I_{L_2} \, (\omega) = \left( \frac{L_2}{L_1} \right)^3 \, I_{L_1} \!\! \left( \omega_0 - (\omega_0 - \omega) \, \left( \frac{L_2}{L_1} \right)^2 \right).
\end{equation}
Here $I_{L_{1,2}} (\omega)$ are the Raman spectra of identical particles with sizes $L_{1,2}$, respectively, which both have the same shape; $L_2$ is
slightly different from $L_1$. Empirically, this EKFG scaling~\eqref{scaling} may be extended onto DMM-BPM approach, as well.

\section{Optical phonon line broadening in disordered nanoparticles. } \label{III}

Now we specify the optical phonon damping parameters $\Gamma_n$ introduced in previous section. The source of phonon scattering is assumed to be the particle intrinsic disorder taken in two modifications: the randomness of atomic masses and the defective interatomic bonds.

The Hamiltonian of elastic medium within the harmonic approximation reads
\begin{equation}\label{H}
  {\cal H} = \sum_l \frac{p^2_l}{2 M_l} +  \frac{1}{2} \, \sum_{l l^\prime} K_{l l^{\prime}} \left( \mathbf{r}_l - \mathbf{r}_{l^\prime} \right)^2,
\end{equation}
where the first sum describes the kinetic energy of atoms with masses $M_l$ packed in a lattice. These atoms are connected by the springs with rigidities $ K_{l l^\prime}$, the elastic energy being proportional to the squared difference of atomic displacements ${\bf r}_l$ from their equilibrium lattice positions ${\bf R}_l$. Firstly, we incorporate the Gaussian weak delta-correlated disorder via the spatial randomness of atomic masses $M_l$ characterized by the mean value $ M  =  \langle \, M_l \, \rangle $ and the  variation $\delta m_l$ with zero average $\langle \, \delta m_l  \rangle =0$ and delta-functional pairwise correlator
\begin{equation}\label{dis1}
\frac {\langle \,\delta m_l \, \delta m_{l^\prime} \, \rangle}{ M^2 }\, = S_m \,\delta_{l l^\prime}.
\end{equation}
The mass disorder term enters Eq.~\eqref{H} as follows:
\begin{equation}\label{H imp 1}
{\cal H}_{imp} = - \, \frac{1}{2 M^2} \sum_{l} \delta m_l \, p_{l}^2.
\end{equation}
Similarly, we consider the Gaussian delta-correlated randomness of interatomic bonds $K_{l l^{\prime} }$ characterizing this kind of disorder by the average rigidity $K = \langle \, K_{l l^{\prime} } \rangle$ and its variation $\delta k_{l l^{\prime}}$ which also has zero average $\langle \delta k_{l l^{\prime}} \rangle =0$ and  delta-functional correlator
\begin{equation} \label{dis2}
  \frac{\langle \delta k_{l_1 l_1^{\prime}} \delta k_{l_2 l_2^{\prime}} \rangle}{K^2} = S_k \delta_{l_1 l_1^{\prime},l_2 l_2^{\prime}},
\end{equation}
providing the contribution to the Hamiltonian of the form
\begin{equation} \label{H imp 2}
 {\cal H}_{imp} = \frac{1}{2} \, \sum_{l l^\prime} \delta k_{l l^{\prime}}  \left( \mathbf{r}_l - \mathbf{r}_{l^\prime} \right)^2
\end{equation}
The quantities $S_m$ and $S_k$ have the meanings of mass and bond disorder strengths, respectively. Since the averages in Eqs.~\eqref{dis1} and \eqref{dis2}
are nonzero only for defective lattice sites (bonds) they are proportional to the (reduced) concentrations of corresponding impurities multiplied by the  squared reduced amplitudes of their correlators (see, e.g., Refs.~\cite{our3,our4} for details).

Next, executing the procedure of second quantization  (i.e., expressing the displacements $\mathbf{r}_l$ and the momenta $\mathbf{p}_l$ in our Hamiltonian via the bosonic creation/annihilation operators $b^\dagger_n \, (b_n)$ and the phonon eigenfunctions $\mathbf{Y}_n(\mathbf{R}_l)$) as follows
\begin{equation}\label{rlQ}
  \mathbf{r}_l = \frac{1}{\sqrt{2 M}} \sum_n \frac{\mathbf{Y}_n(\mathbf{R}_l)}{\sqrt{\omega_n}} (b_n + b^\dagger_n)
\end{equation}
and
\begin{equation}\label{plQ}
  \mathbf{p}_l = \frac{i \sqrt{M}}{\sqrt{2}} \sum_n \mathbf{Y}_n(\mathbf{R}_l) \sqrt{\omega_n} (b^\dagger_n - b_n),
\end{equation}
we arrive to the Hamiltonian in the form $\mathcal{H}= \mathcal{H}_0 + \mathcal{H}_{imp}$, where the first term stands for the gas of free phonons:
\begin{equation} \label{h0}
  \mathcal{H}_0 = \sum_n \omega_n (b^\dagger_n b_n + 1/2),
  \end{equation}
  and the second one describes the optical phonon scattering by the impurities
\begin{eqnarray} \label{himp1}
  \mathcal{H}_{imp} &=&  \frac{1}{4} \sum_{n,n^\prime, l}  \frac{\delta m_l }{M} \sqrt{\omega_n \omega_{n^\prime}} (b_n +b^\dagger_n)(b_{n^\prime}+b^\dagger_{n^\prime}) \\ && \mathbf{Y}_n(\mathbf{R}_l) \cdot \mathbf{Y}_{n^\prime}(\mathbf{R}_l) \nn
\end{eqnarray}
for random masses and
\begin{eqnarray} \label{himp2}
  \mathcal{H}_{imp} &=& \frac{1}{4M} \sum_{n,n^\prime, \langle l l^\prime \rangle} \frac{\delta K_{l l^\prime}}{\sqrt{\omega_n \omega_{n^\prime}}} (b_n +b^\dagger_n)(b_{n^\prime}+b^\dagger_{n^\prime}) \\ && (\mathbf{Y}_n(\mathbf{R}_l)-\mathbf{Y}_n(\mathbf{R}_{l^\prime})) \cdot (\mathbf{Y}_{n^\prime}(\mathbf{R}_l)-\mathbf{Y}_{n^\prime}(\mathbf{R}_{l^\prime})). \nonumber
\end{eqnarray}
for random bonds.
It seems that the scattering terms \eqref{himp1} and \eqref{himp2} principally differ one from another: the first expression represents the scattering on site disorder while the second one corresponds to scatterers in the form of a bond disorder. It is really essential for acoustic phonons, where the bond disorder scattering amplitude contains an
extra power of momentum due to this fact and therefore the phonon lifetimes evaluated for these two mechanisms will have different momentum (and therefore particle size) dependencies. At the same time, for optical phonons and neighboring atoms we get
$\mathbf{Y}_n(\mathbf{R}_l) \approx -\mathbf{Y}_n(\mathbf{R}_{l^\prime})$, and Eq.~\eqref{himp2} becomes indistinguishable from Eq.~\eqref{himp1}, up to the replacement of the prefactor. Thus, we conclude that the mass disorder and the bond disorder yield similar contributions, so we can just investigate one of them (say, mass disorder) and then replace in final formulas $S_m \to S = S_m + S_k$.

We formulate the disordered diagram technique for operators $\phi_n = b_n +b^\dagger_n $ and Green's functions $-i \langle \hat{T} \phi_n \phi_n \rangle$ (here $\hat{T}$ is the time-ordering operator). Upon the disorder averaging the latter obtains the form
\begin{equation}\label{DQ}
D_n(\omega)= \frac{2  \omega_n}{\omega^2 - \omega^2_n - 2 \omega_n \Pi_n (\omega)},
\end{equation}
where the self-energy part $\Pi_n (\omega)$ could be calculated using different approximation schemes (see Ref.~\cite{our3,ourK} for details). Here we just present the results.

\noindent
{\it Dilute weak (Born) disorder.}
When the phonon levels are separated (i.e., not overlapped due to the phonon scattering by disorder) one should use the self consistent Born approximation which yields
\begin{equation}\label{GS}
  \Gamma_n (L, S, p) =  \omega_n \, \mu_n (p) \, \sqrt{S} \left( \frac{a_0}{L} \right)^{3/2}.
\end{equation}
Here $\mu_n (p) $  is certain (known) quantum number and shape dependent factor which could be calculated analytically for cubic, spherical, and cylindrical particles and numerically for other particle shapes.

For overlapped levels we utilize the fact that the phonon eigenfunctions are not essentially different from the plane waves in this case. It gives
\begin{equation}\label{GO2}
\Gamma_n (L, S,  p) = \omega_{n} \, \nu_n (p) \, S \, \frac{a_0}{L},
\end{equation}
$\nu_n (p)$ being another known $n$ and $p$ dependent factor.

The crossover scales between these two regimes could be calculated  as a crossover disorder $S_c \sim a_0 / L $ (at fixed particle size), and {\it vice versa}, $L_c \sim a_0 /S$; for shape dependent prefactors see, e.g., Refs.~\cite{our3,our4}.

The proper formalism for the {\it dilute strong (binary) disorder} that allows to evaluate the phonon lifetimes and the energy of localized phonon-impurity bound state is the T-matrix approximation. Within this formalism the intensity of  impurity scattering is governed by the effective impurity potential defined as
\begin{equation}\label{U m}
U_m = \frac{\delta m }{M + \delta m}
\end{equation}
for mass disorder and as
\begin{equation} \label{U_k}
U_k = \frac{\delta k}{K}
\end{equation}
for bond disorder.
The resulting (resonant) enhancement of damping obtains the form \cite{our3,our4}
\begin{equation} \label{Gamma strong}
\Gamma_n = \omega_n \, 4 \pi F \, c_{imp} \frac{\xi}{a_0} \, \frac{q_n \xi}{1 + (q_n \xi)^2},
\end{equation}
where $c_{imp}$ is the impurity concentration, and the large spatial scale $\xi \gg  a_0$ measures the proximity to the limiting impurity potential $U_{min}$.
Remind that shape, quantum number, and size dependencies of the damping contain in the quasimomentum $q_n$.

The energy of the phonon-impurity localized bound state lies slightly above the maximal phonon frequency
\begin{equation} \label{omega loc}
\frac{\omega_{loc} - \omega_0}{\omega_0} = F \, \left( \frac{a_0}{\xi} \right)^2 .
 \end{equation}

\section{Numerical approach} \label{IV}

We support our analytical findings by proper numerics demonstrating the equivalence of analytical and numerical results and thus proving the reliability of our approaches. We  employ the exact diagonalization of the dynamical matrix with the Gaussian disorder. The eigenmodes $| \varepsilon \rangle$ obtained in that way are utilized when calculating
the broadening for the $n$-th mode of a pure particle. Averaging over disorder is realized with the use of the formula:
\begin{equation}
  \overline{\sum_\varepsilon \delta(\omega- \varepsilon) | \langle n | \varepsilon \rangle|^2},
\end{equation}
where the overline stands for averaging. We averaged over several hundreds configurations for each particle size/disorder strength.

As usually in disordered systems, the solution of  eigenproblem for any particular realization of disorder yields real (albeit, disorder modified) eigenfrequencies and eigenfunctions. The broadening arises upon the averaging due to non-equivalence of disorder realizations; for particles, even the number of impurities in a particle fluctuates around its mean (over the ensemble) value. We fit the (broadened) spectral lines by the Lorentzians.

\begin{figure}[t]
  \centering
  % Requires \usepackage{graphicx}
  \includegraphics[width=8.cm]{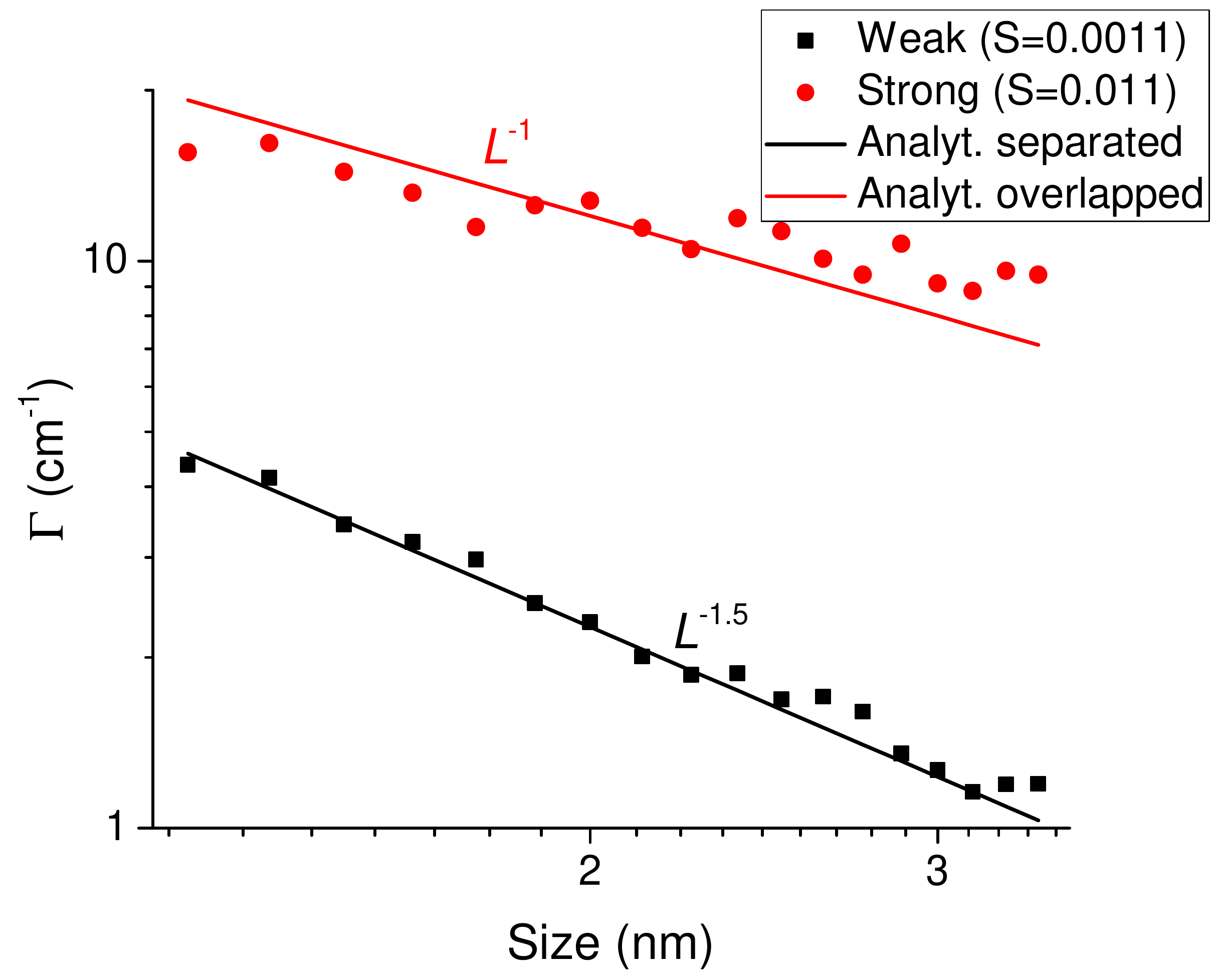}
  \caption{The phonon linewidth $\Gamma_1$ in a spherical nanodiamond plotted as a function of the particle size $L$ for two disorder strengths $S_m=0.0011$ (black squares) and $S_m=0.011$ (red dots). The range of $L$ shown in this Figure corresponds to the regime of separated levels in the former case and to the regime of overlapped levels in the latter one. The lines represent the predictions of analytical theory. }
\label{FigL}
\end{figure}

\begin{figure}[t]
  \centering
  % Requires \usepackage{graphicx}
  \includegraphics[width=8.cm]{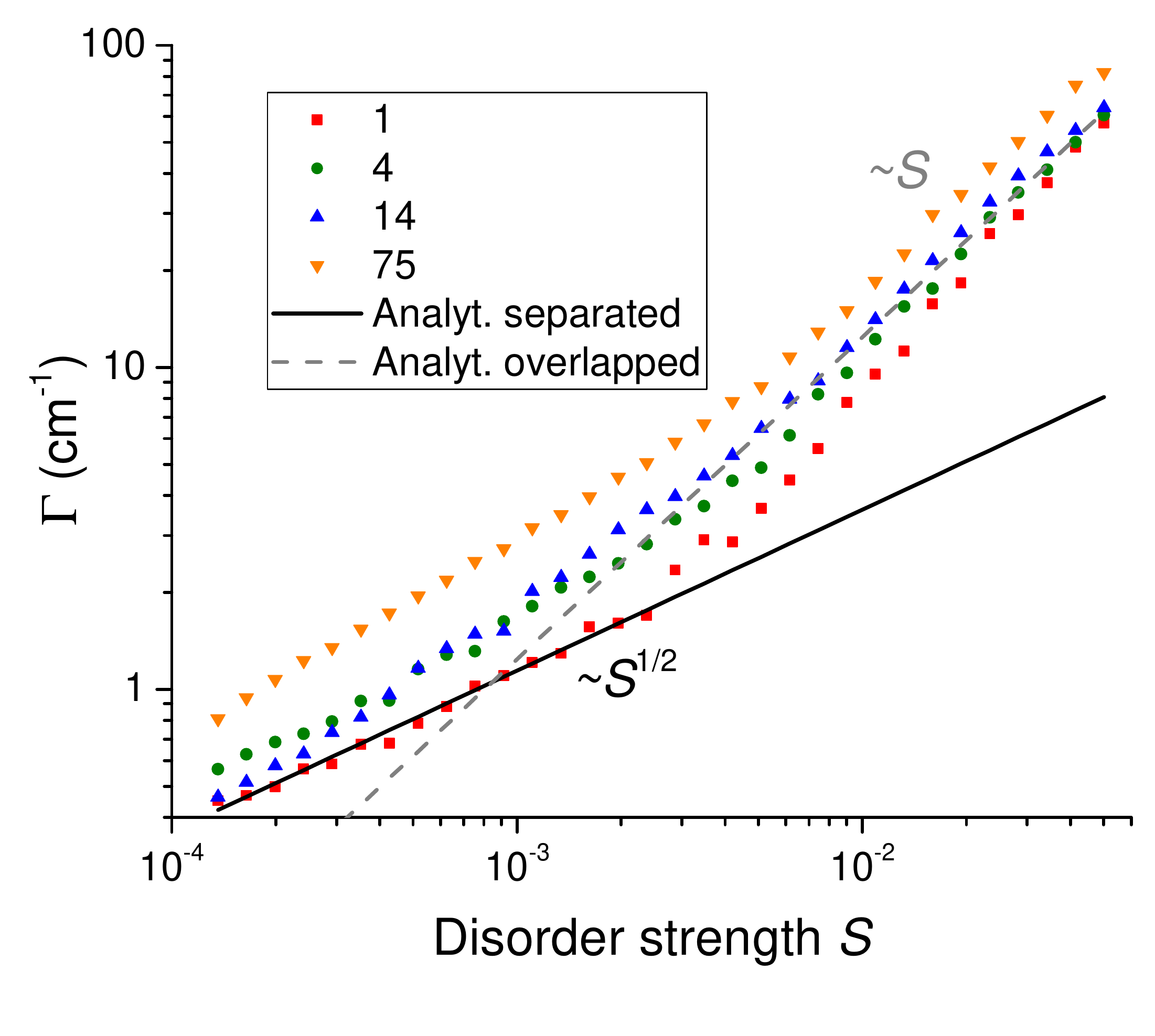}
  \caption{The phonon linewidth $\Gamma_n$ plotted as a function of disorder strength $S_m$ for several phonon eigenmodes ($n=1, 4, 14, 75$) of a spherical $3 \, \text{nm}$ nanodiamond. Solid and dashed lines show the analytical predictions for the first mode in separated and overlapped regimes, respectively. The highest phonon modes have broader lines. Both the functional dependencies and the crossovers are clearly seen.}
\label{FigS}
\end{figure}

In order to be specific we check numerically the case of mass disorder (the results for random bonds is similar, see Ref.~\cite{ourK}).
In Fig.~\ref{FigL} we depict the comparison of our numerical and analytical results for the phonon linewidth as a function of particle size $L$ for both regimes of separated and overlapped levels whereas Fig.~\ref{FigS} demonstrates the linewidth dependence on the disorder strength $S_m$. We emphasize a very good agreement between the analytics and the numerics including the prefactors, functional dependencies and crossovers visible in these Figures.

Now we are ready to apply our theory for the detailed analysis of experimental data.

\section{Analysis of experimental data} \label{V}

The main goal of the present paper is to promote the approach formulated in Refs.~\cite{ourDMM,ourEKFG,our3,our4,ourK} as a fruitful tool of analysis of experimental data on the Raman scattering in nanopowders of nonpolar crystals, and to demonstrate its capabilities. Before we start to interpret the particular experiments, let us present two important general results of this approach.

\noindent
(A) It is mentioned in the great number of experimental papers that the width of the Raman peak in {\it nanopowders} is by the order of magnitude bigger than that  in the corresponding bulk materials. For instance,  this width in nanodiamonds is $\sim 10 \, cm^{-1}$ whereas in large diamond crystals it is $\sim 1 \, cm^{-1}$. This increase of the width finds its naturally explanation in our theory which predicts the phonon linewidths inversely proportional to some powers of the particle size $\Gamma \propto L^{-x}$, with $x=1$ or $3/2$, depending on the regime of scattering. Simple estimates reveal that even the mass disorder alone taken in the form of the widespread in nanodiamonds NV  (nitrogen+vacancy) centers is capable to provide the experimentally measured values of the Raman peak width, and the required concentration of centers $c_{imp} \sim 1-3 \% $ agrees with the typical values of a carbon admixture in diamonds detected chemically.

\noindent
(B) The analysis of experimental data on the Raman scattering in nanodiamonds undertaken in Ref.~\cite{yoshikawa1993raman} revealed roughly the inverse  dependence of the Raman peak width in nanodiamonds on their size $L$. Our analysis of experimental conditions described in Ref.~\cite{yoshikawa1993raman} allows us to conclude that the particles investigated were relatively large, thus belonging to the overlapped regime of impurity scattering. It is precisely the regime where our theory predicts the phonon linewidths of the form $\Gamma \propto 1/L$.

The theory of the phonon line broadening in {\it weakly disordered} nanoparticles presented in this paper is a part of more general approach of Refs.~\cite{ourDMM,ourEKFG,our3,our4,ourK} accounting for also strong disorder, surface corrugations, particle coating and amorphization, etc. It yields a possibility to build up the regular method of analyzing the Raman spectra of nanopowders of nonpolar crystals more reliably and precisely than the previously used ones. Below we demonstrate that this method allows us not only to fit the experimental curves but also to extract from the data four parameters important for the nanopowder specification, namely (i) the mean particle size in a powder $L$, (ii) the standard deviation of size distribution function $\delta L$, (iii) the disorder strength parameter $S$, and, with less accuracy, (iv) the particle shape parameterized by the faceting number $p$. It becomes possible only with the microscopic theory~\cite{our3,our4,ourK} of the phonon line broadening at hands. Indeed, this theory predicts various broadenings $\Gamma_n$ for various phonon modes $\omega_n$. While the empirical theory of Refs.~\cite{ourDMM,ourEKFG} which utilizes the single broadening parameter $\Gamma$ for all phonon modes (it allows to determine the abovementioned parameters with one significant digit) the microscopic theory improves the accuracy to two digits. Moreover, it makes possible to determine from the experiment the particle shape,
which is the more delicate effect $\sim 10-20 \%$ of the total measured entities.

The thorough analysis of a Raman experiment should include the effects of asymmetry of phonon lines and the more detailed discussion of all possible sources of broadening. In the present paper, we undertake the simplified version of such analysis examining the experimental data of Refs.~\cite{yoshikawa1995raman,shenderova2011nitrogen,gao2017origin} with the use of the EFKG method. Now let us present some details of our fit of the experiments.

Performing the program outlined in previous sections in practice we restrict the sum in $n$ in Eq.~\eqref{I1} counting only the levels with $I_n /I_1 > 0.01$. The broadening parameter $\Gamma_n$ should be taken either in the form~\eqref{GS} or as ~\eqref{GO2}, the choice must be justified after the fit. We notice that the complicated regime where for small particles the levels are separated while for large ones they are are overlapped sometimes emerges. For experimentally relevant case the levels are overlapped, and instead of~\eqref{GO2} one can use
\begin{eqnarray} \label{damp1}
% \nonumber to remove numbering (before each equation)
  \Gamma_1(L,S,p) &=& g(S,p) \frac{a}{L}, \\
  \Gamma_n(L,S,p) &=& g(S,p)\frac{a}{L} \sqrt{\frac{\omega_0 - \omega_n(p,L_0)}{\omega_0 - \omega_1(p,L_0)}}. \nonumber
\end{eqnarray}
Here we introduce the characteristic broadening $g(S,p)$. In the last equation we utilized the relation $\Gamma_n \propto q_n$ peculiar for the overlapped regime.

Before the fitting of experimental data one should choose the reasonable distribution function $f$, e.g., the log-normal or the Gaussian one, with free fitting parameters. The log-normal distribution seems to us (and is proven in practice) the more appropriate one at least for for nanodiamonds due to its effective cutting of at very small particles and due to its long tailing for the big ones. We write:
\begin{equation}\label{log-normal}
  f(L,\mu,\sigma) = \frac{1}{L \sigma \sqrt{2 \pi}} e^{-\frac{(\ln{L}-\mu)^2}{2\sigma^2}};
\end{equation}
its discretized version appropriate for Eq.~\eqref{I1} reads:
\begin{eqnarray}\label{log-normal2}
  f(L,\mu,\sigma) \rightarrow f(L_i,\mu,\sigma), \\ L_i = e^{\mu + \sigma^2/2} + \frac{i\sigma}{2}, \, i=-N ... N, \nonumber
\end{eqnarray}
with proper $N$  (say, $N=7$). Importantly,  $e^{\mu + \sigma^2/2}$  is the mean particle size. The generalization of Eqs.~\eqref{log-normal},  \eqref{log-normal2} for any other distribution function is straightforward.

The fit problem is formulated as follows. Let $(\omega_k, I_{exp}(\omega_k))$ be the experimentally measured points of the powder Raman spectrum in the total amount of $J$ points. If we  denote the scattering intensity, which takes into account Eqs.~\eqref{damp1} and \eqref{log-normal2}, as $I(\omega, g, p, \mu, \sigma)$ we can determine the parameters $g, \mu, \sigma$ for each shape $p$ using the least squares method:
\begin{equation}\label{chi1}
  \chi^2_p = \frac{1}{K} \sum_k |I_{exp}(\omega_k) - I(\omega_k, g, p, \mu, \sigma)|^2 \rightarrow \mathrm{min}.
\end{equation}
The smallest $\chi^2_p$ defines the particle shape, $\mu$ and $\sigma$ determine the mean particle size $L= e^{\mu + \sigma^2/2}$ and the standard deviation $\delta L = e^{\mu + \sigma^2/2}\sqrt{\left( e^{\sigma^2} - 1\right)}$, respectively. The disorder strength $S$ is extracted from $g(S,p)$.

When testing the nanodiamonds, we use the values of Keating parameters $A = 1193.75$ cm$^{-1}$ and  $B = 139.25 $ cm$^{-1}$, see Ref.~\cite{yoshikawa1995raman}.

\begin{figure}[t]
  \centering
  % Requires \usepackage{graphicx}
  \includegraphics[width=8.cm]{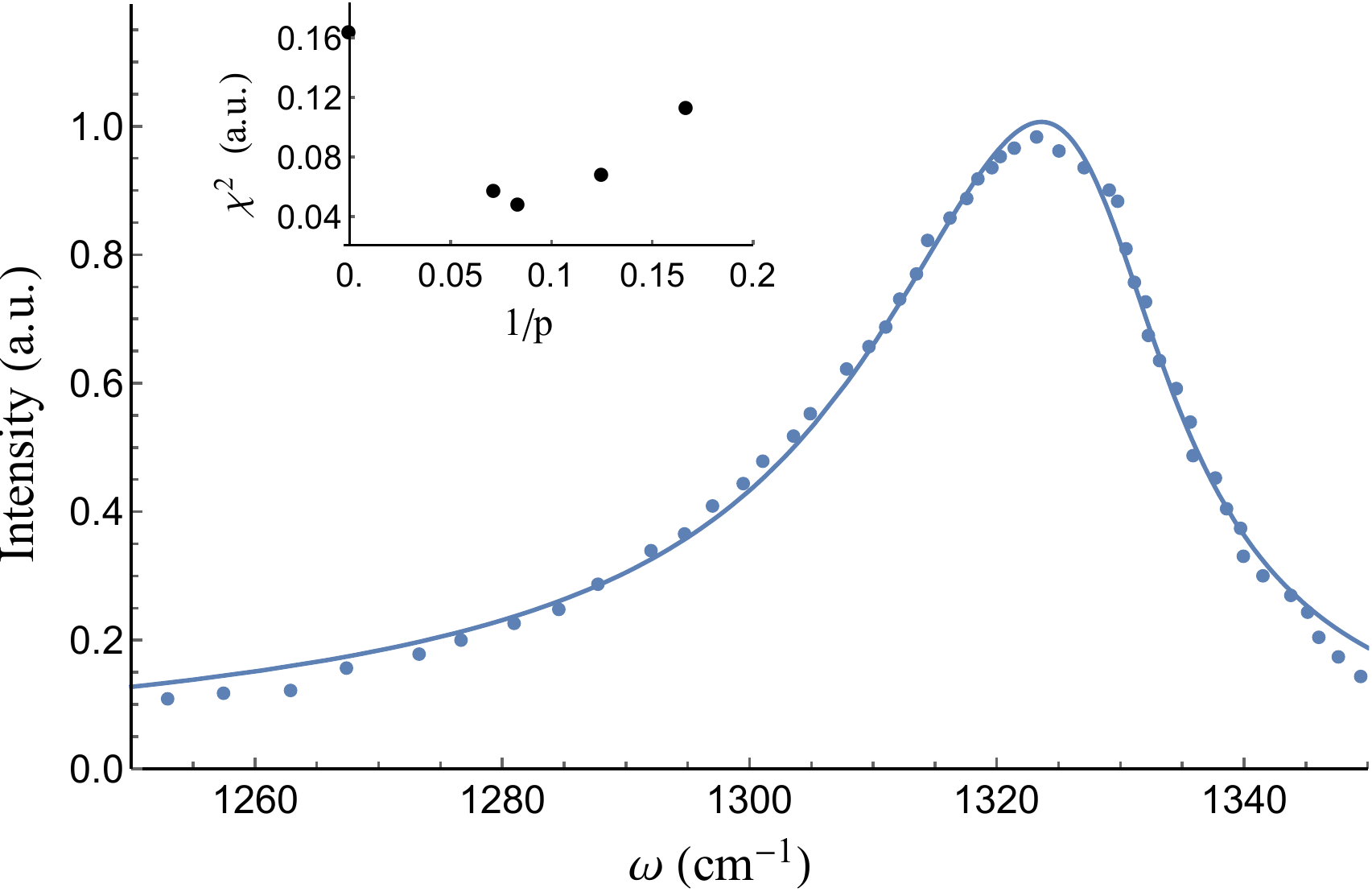}
  \caption{The Raman spectrum of a diamond nanopowder from Ref.~\cite{yoshikawa1995raman} (points) and its fit with the use of our theory (solid line). Inset demonstrates the values of $\chi^2$ parameter obtained for the fits with the use of various polyhedra. The minimum is around $p = 12$
\label{FigFitY}}
\end{figure}

\noindent
(C) The first (nanodiamonds) data to be re-examined with the use of our theory have been reported by Yoshikawa and co-authors.~\cite{yoshikawa1995raman}
The plot is shown in Fig.~\ref{FigFitY}. The best fit has been obtained for the log-normal distribution and for  the broadening which occurs in the overlapped regime. It yields $L \approx 2.4 \, \text{nm}$, $\delta L \approx 0.6 \, \text{nm}$, and $S \approx 0.03$. The particles shape was found to be a dodecahedron with faceting number $p =12$, however, the truncated octahedra yields quite close value of $\chi^2$ (see inset).

\begin{figure}[t]
  \centering
  % Requires \usepackage{graphicx}
  \includegraphics[width=8.cm]{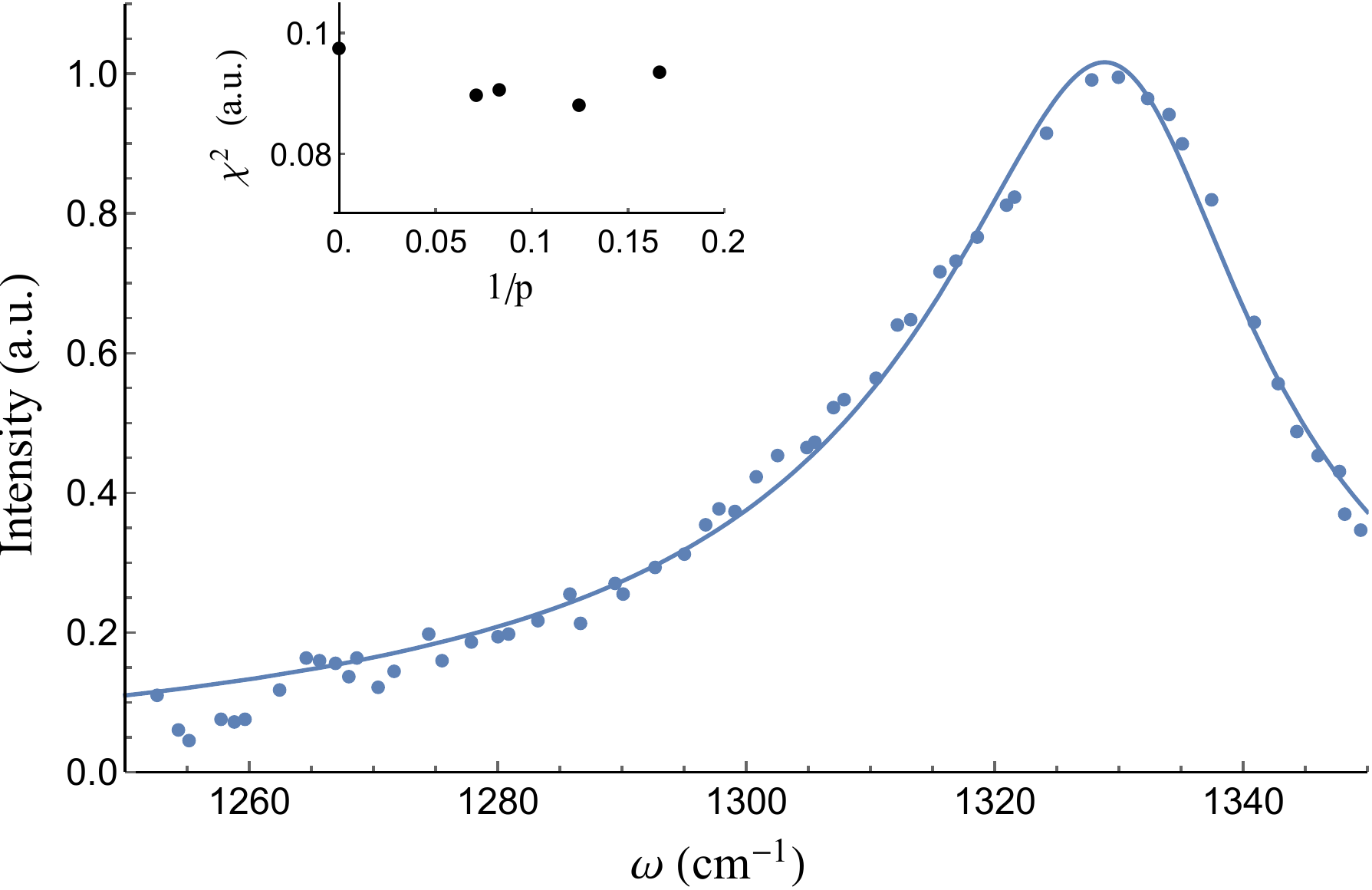}
  \caption{The same as in Fig.~3 taken from Ref.~\cite{shenderova2011nitrogen}. The minimum around $p = 8$ in the inset is more flat.
\label{FigFitS}}
\end{figure}

\noindent
(D) Second, we address the experimental data on nanodiamonds presented by Shenderova group in Ref.~\cite{shenderova2011nitrogen}. The results are shown in Fig.~\ref{FigFitS}. Once again, the best fit was obtained for the log-normal distribution and the overlapped regime of broadening. The obtained powder parameters are $L \approx 3.5 \, \text{nm}$, $\delta L \approx 1.3 \, \text{nm}$, and $S \approx 0.06$. Concerning the particle shape we see, that truncated octahedra ($p=14$), dodecahedra ($p=12$), and octahedra ($p=8$) yield very similar results (the minimum in the inset is more flat than for previous data). Formally, we should end up with the value $p=8$; however, comparing this data with the results of (C) we believe that to rough the accuracy is the more appropriate option in this situation.  It yields $p \sim 10$ from the results of two groups.

\begin{figure}[t]
  \centering
  % Requires \usepackage{graphicx}
  \includegraphics[width=8.cm]{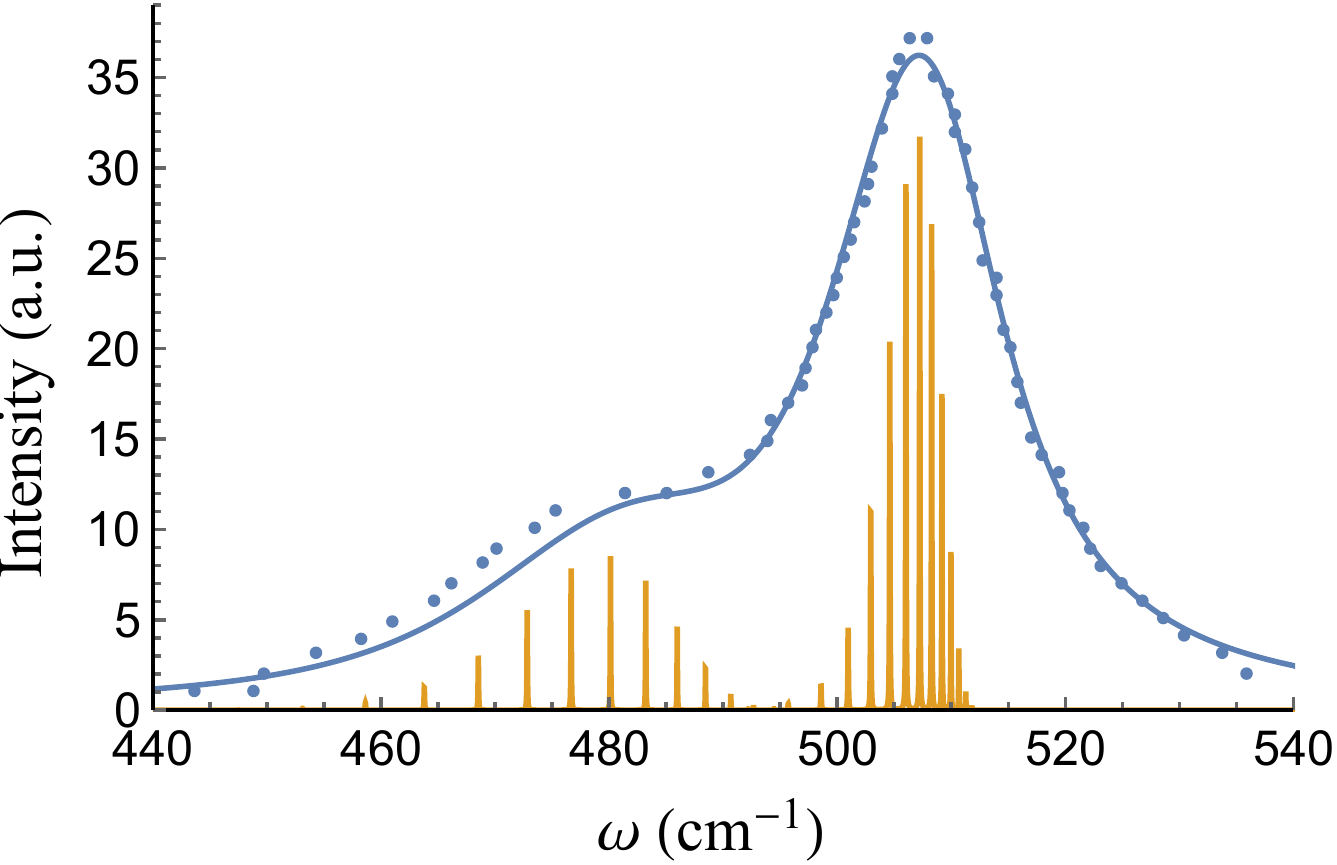}
  \caption{The Raman spectrum of a crystalline {\rm Si} nanopowder from Ref.~\cite{gao2017origin} (points) and its fit with the use of our theory (solid line). The two-peak structure is reproduce without complication of the model. Note that the continuous EKFG approach is applicable only qualitatively deeply on the left shoulder
  where the fit quality is seen to be worse. The orange comb-like lines show the mode structure for the superposition of all discrete 15 sizes in the size distribution. The groups of peaks corresponding to the first and to the second Raman active modes are well visible.
\label{FigFitSi}}
\end{figure}

\noindent
(E)
Finally, we elaborate the experimental data for {\rm Si} nanocrystals obtained by Gao and Yin in Ref.~\cite{gao2017origin}. In this paper, the authors presented the Raman data for different particle sizes. The data reveal rather large Raman shift for reported particle sizes, if the conventional Keating parameters $A \approx 500$ cm$^{-1}$ and $B \approx 21$ cm$^{-1}$ are used. Using the quantization rule of Ref.~\cite{ourDMM}, we find that the values $A \approx 477.8$ cm$^{-1}$ and $B \approx 40.5$ cm$^{-1}$ are more appropriate to fit the data. Next, we examine the Raman spectrum for particles with the size $2.7$ nm within the framework of our scheme. Despite of the fact that the size of these particles is only about 5 bigger than the lattice parameter for {\rm Si}, the EKFG fit shown in Fig.~\ref{FigFitSi} works very well. Here we use the Gaussian distribution function, spherical particles, and the broadening in separated regime ($\Gamma \approx 18$ cm$^{-1}$). We find the mean particle size $L \approx 2.1$ nm and the distribution function variance $\delta L \approx 0.2$ nm. Within our approach the resulting Raman spectrum consists of two contributions: the first Raman-active mode centered at $\omega \approx 505$ cm$^{-1}$ and the second Raman-active mode (the reminder of the first Raman-active band shrunk in the crude EKFG approach into the single line) at $\omega \approx 481$ cm$^{-1}$, which provides the feature at the left shoulder of the spectrum. We would like to point out that this feature is very typical and always seen for narrow size distribution functions (see, e.g., Ref.~\cite{our3}); it is usually smeared out by the wide size distribution in nanopowders. The authors of Ref.~\cite{gao2017origin} undertook special technical efforts in order to make the size distribution function narrower; however, they used the more involved model to describe the two-peak structure of the spectrum. Our approach provides an alternative explanation of these data which is free of unnecessary complications.  Finally, Fig.~\ref{FigFitSi} illustrates that broad particle size distribution can not explain total broadening and shape of Raman peak for nanoparticles, because the signal is still strong at the frequencies larger than 520 cm$^{-1}$. The contributions to the peak broadening due to size dispersion and disorder act simultaneously.

\section{Discussion and conclusions} \label{VI}

We believe that our method could be very useful for the analysis of Raman spectra of nanopowders of nonpolar crystals. However, it is not completely free of some
disadvantages (or peculiarities) which we would like to mention now.

First, it has been shown in Ref.~\cite{ourK} that for two-component particles which consist of the relatively clean crystalline core and the strongly disordered
(even amorphous) surface shell the ``particle size'' extracted from the data with the use of our method will correspond to the core size rather to the entire particle size including its coating.
This type of particles is very widespread; in order to analyse them, one should properly modify our approach or accompany it by some extra measurements.

Second, the disorder strength parameter $S$ extracted from our analysis is in fact a sum of several contributions ($S = S_m + S_k$ in the simplest case).
In is not known {\it a priori} in what proportion the randomness of interatomic bonds and the mass disorder contribute to $S$. The simple general arguments based on the virial theorem and claiming that the energy associated with the randomness should be equally distributed among the kinetic term (mass disorder) and the potential term (bond disorder) are not always applicable.

Moreover, even for weak (mass) disorder the strength  parameter $S_m= c_{m, imp} (\delta m /M)^2$ ( $S_k=c_{k, imp} (\delta k /K)^2$ for bond disorder) extractable from our data analysis is a product of two characteristics, each of them being of particular interest {\it per se}. Therefore, for the comprehensive characterization of a nanopowder our theory should be supplemented by the proper analysis of its chemical composition.

Some conclusions, however, could be drawn right now. The obtained value $S \approx 0.03$ is too large to be explained by the isotopic disorder: for $c_{imp} \sim 10^{-2} \, (1 \%)$ and $\delta m/M  = 1/12$ one obtains $S \sim 10^{-4} $. Furthermore,
it is evident that for  reasonable concentration of impurities $c_{imp} \sim 10^{-3} - 10^{-1}$ only the strong scatterers could provide the desired value of $S $. There are two following attractive candidates for this role in diamonds, the famous NV (nitrogen + vacancy) centers and the silicon + vacancy complexes, both of them include the vacancy which in our theory works as an infinite repulsive on-site potential. The typical concentration of nitrogen in diamonds is $c_{imp} \sim 1-3 \%$; the estimated value of $U_m$ for NV centers (NV centers are strong scatterers) is capable to provide   $S \approx 0.03 $; however, this issue deserves a more detailed treatment.

Our theory deals with the elastic processes of phonon scattering by disorder, and, therefore, with Raman spectra at low temperatures. It is interesting to investigate this issue at higher temperatures when the inelastic processes of phonon scattering by each other could modify the situation. Here we just mention the intriguing possibility to observe the localization-delocalization temperature crossover induced by the many-body localization effects predicted in Refs.~\cite{altshuler1997,gorniy2017} for the electron counterpart (interacting electrons in a quantum dot) of present problem.

To conclude, we analyzed the experimental Raman spectra of nanopowders of nonpolar crystals utilizing our theory of Raman scattering in disordered particles developed earlier. We explained with the use of our theory the large width of the Raman peak in nanoparticles and its inverse dependence on the particle size observed experimentally. We also demonstrated that this theory allows us to extract confidently from the Raman data three important microscopic parameters such as the mean particle size, the variance of the particle size distribution function, the strength of intrinsic disorder, and to  estimate the effective faceting number parameterizing the particle shape.

\section*{Acknowledgments}

The authors are thankful to Igor Gornyi for valuable comments. This work is supported by the Russian Science Foundation (Grant No. 19-72-00031).

\section*{Data availability}

The data supporting our findings are available from the corresponding authors upon reasonable request.

\bibliography{KFG}

\end{document}